\documentclass[twocolumn,pre]{revtex4}

\usepackage{dcolumn}
\usepackage{amsmath}

\usepackage{graphicx}

\begin{document}

\newcommand{\Ord}{{\rm O}}
\newcommand{\eref}[1]{(\ref{#1})}
\newcommand{\etal}{{\it{}et~al.}}

\newlength{\figurewidth}
\ifdim\columnwidth<10.5cm
  \setlength{\figurewidth}{0.95\columnwidth}
\else
  \setlength{\figurewidth}{10cm}
\fi
\setlength{\parskip}{0pt}
\setlength{\tabcolsep}{6pt}

\title{Fast algorithm for detecting community structure in networks}
\author{M. E. J. Newman}
\affiliation{Department of Physics and Center for the Study of Complex
Systems, University of Michigan, Ann Arbor, MI 48109--1120}
\begin{abstract}
It has been found that many networks display community structure---groups
of vertices within which connections are dense but between which they are
sparser---and highly sensitive computer algorithms have in recent years
been developed for detecting such structure.  These algorithms however are
computationally demanding, which limits their application to small
networks.  Here we describe a new algorithm which gives excellent results
when tested on both computer-generated and real-world networks and is much
faster, typically thousands of times faster than previous algorithms.  We
give several example applications, including one to a collaboration network
of more than $50\,000$ physicists.
\end{abstract}
\maketitle

\section{Introduction}
There has in recent years been a surge of interest within the physics
community in the properties of networks of many kinds, including the
Internet, the world wide web, citation networks, transportation networks,
software call graphs, email networks, food webs, and social and biochemical
networks~\cite{Strogatz01,AB02,DM03b,Newman03d}.  One property that has
attracted particular attention is that of ``community structure'': the
vertices in networks are often found to cluster into tightly-knit groups
with a high density of within-group edges and a lower density of
between-group edges.  Girvan and Newman~\cite{GN02,NG04} proposed a
computer algorithm based on the iterative removal of edges with high
``betweenness'' scores that appears to identify such structure with some
sensitivity, and this algorithm has been employed by a number of authors in
the study of such diverse systems as networks of email messages, social
networks of animals, collaborations of jazz musicians, metabolic networks,
and gene networks~\cite{GN02,NG04,WH03,HHJ03,Guimera03,TWH03,GD04}.  As
pointed out by Newman and Girvan~\cite{NG04}, the principle disadvantage of
their algorithm is the high computational demands it makes.  In its
simplest and fastest form it runs in worst-case time $\Ord(m^2n)$ on a
network with $m$ edges and $n$ vertices, or $\Ord(n^3)$ on a sparse graph
(one for which $m$ scales with $n$ in the limit of large~$n$, which covers
essentially all networks of current scientific interest, with the possible
exception of food webs).  With typical computer resources available at the
time of writing, this limits the algorithm's use to networks of a few
thousand vertices at most, and substantially less than this for interactive
applications.  Increasingly however, there is interest in the study of much
larger networks; citation and collaboration networks can contain millions
of vertices~\cite{Redner98,Newman01a} for example, while the world wide web
numbers in the billions~\cite{KL01}.

In this paper, therefore, we propose a new algorithm for detecting
community structure.  The algorithm operates on different principles to
that of Girvan and Newman~(GN) but, as we will show, gives qualitatively
similar results.  The worst-case running time of the algorithm is
$\Ord((m+n)n)$, or $\Ord(n^2)$ on a sparse graph.  In practice, it runs to
completion on current computers in reasonable times for networks of up to a
million or so vertices, bringing within reach the study of communities in
many systems that would previously have been considered intractable.

\section{The algorithm}
Our algorithm is based on the idea of modularity.  Given any network, the
GN community structure algorithm always produces \emph{some} division of
the vertices into communities, regardless of whether the network has any
natural such division.  To test whether a particular division is meaningful
we define a quality function or ``modularity''~$Q$ as follows~\cite{NG04}.
Let $e_{ij}$ be the fraction of edges in the network that connect vertices
in group~$i$ to those in group~$j$, and let $a_i=\sum_j
e_{ij}$~\footnote{As discussed in~\cite{NG04}, each edge should contribute
only to $e_{ij}$ once, either above or below the diagonal, but not both.
Alternatively, and more elegantly, one can split the contribution of each
edge half-and-half between $e_{ij}$ and $e_{ji}$, except for those edges
that join a group to itself, whose contribution belongs entirely to the
single diagonal element $e_{ii}$ for the group in question.}.  Then
\begin{equation}
Q = \sum_i (e_{ii} - a_i^2)
\end{equation}
is the fraction of edges that fall within communities, minus the expected
value of the same quantity if edges fall at random without regard for the
community structure.  If a particular division gives no more
within-community edges than would be expected by random chance we will get
$Q=0$.  Values other than~0 indicate deviations from randomness, and in
practice values greater than about $0.3$ appear to indicate significant
community structure.  A number of examples are given in
Ref.~\onlinecite{NG04}.

But this now suggests an alternative approach to finding community
structure.  If a high value of $Q$ represents a good community division,
why not simply optimize~$Q$ over all possible divisions to find the best
one?  By doing this, we can avoid the iterative removal of edges and cut
straight to the chase.  The problem is that true optimization of~$Q$ is
very costly.  The number of ways to divide $n$ vertices into $g$ non-empty
groups is given by the Stirling number of the second kind~$S_n^{(g)}$, and
hence the number of distinct community divisions is $\sum_{g=1}^n
S_n^{(g)}$.  This sum is not known in closed form, but we observe that
$S_n^{(1)}+S_n^{(2)}=2^{n-1}$ for all~$n>1$, so that the sum must increase
at least exponentially in~$n$.  To carry out an exhaustive search of all
possible divisions for the optimal value of~$Q$ would therefore take at
least an exponential amount of time, and is in practice infeasible for
systems larger than twenty or thirty vertices.  Various approximate
optimization methods are available: simulated annealing, genetic
algorithms, and so forth.  Here we consider a scheme based on a standard
``greedy'' optimization algorithm, which appears to perform well.

Our algorithm falls in the general category of agglomerative hierarchical
clustering methods~\cite{Everitt74,Scott00}.  Starting with a state in
which each vertex is the sole member of one of $n$ communities, we
repeatedly join communities together in pairs, choosing at each step the
join that results in the greatest increase (or smallest decrease) in~$Q$.
The progress of the algorithm can be represented as a ``dendrogram,'' a
tree that shows the order of the joins (see Fig.~\ref{zachary}, for an
example).  Cuts through this dendrogram at different levels give divisions
of the network into larger or smaller numbers of communities and, as with
the GN algorithm, we can select the best cut by looking for the maximal
value of~$Q$.

Since the joining of a pair of communities between which there are no edges
at all can never result in an increase in~$Q$, we need only consider those
pairs between which there are edges, of which there will at any time be at
most~$m$, where $m$ is again the number of edges in the graph.  The change
in~$Q$ upon joining two communities is given by $\Delta Q = e_{ij} + e_{ji}
- 2a_ia_j = 2(e_{ij} - a_ia_j)$, which can clearly be calculated in
constant time.  Following a join, some of the matrix elements $e_{ij}$ must
be updated by adding together the rows and columns corresponding to the
joined communities, which takes worst-case time~$\Ord(n)$.  Thus each step
of the algorithm takes worst-case time $\Ord(m+n)$.  There are a maximum of
$n-1$ join operations necessary to construct the complete dendrogram and
hence the entire algorithm runs in time $\Ord((m+n)n)$, or $\Ord(n^2)$ on a
sparse graph.  The algorithm has the added advantage of calculating the
value of~$Q$ as it goes along, making it especially simple to find the
optimal community structure.

It is worth noting that our algorithm can be generalized trivially to
weighted networks in which each edge has a numeric strength associated with
it, by making the initial values of the matrix elements~$e_{ij}$ equal to
those strengths, rather than just zero or one; otherwise the algorithm is
as above and has the same running time.  The networks studied in this paper
however are all unweighted.

\section{Applications}
\begin{figure}
\begin{center}
\resizebox{\figurewidth}{!}{\includegraphics{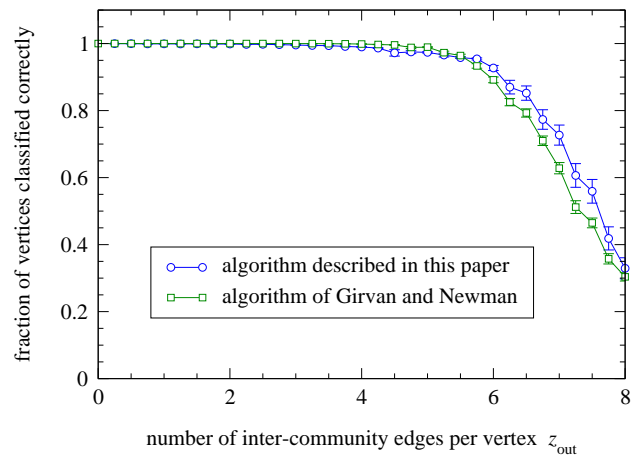}}
\end{center}
\caption{The fraction of vertices correctly identified by our algorithms in
the computer-generated graphs described in the text.  The two curves show
results for the new algorithm (circles) and for the algorithm of Girvan and
Newman~\cite{GN02} (squares).  Each point is an average over 100 graphs.}
\label{correct}
\end{figure}

As a first example of the working of our algorithm, we have generated using
a computer a large number of random graphs with known community structure,
which we then run through the algorithm to quantify its performance.  Each
graph consists of $n=128$ vertices divided into four groups of~$32$.  Each
vertex has on average $z_\mathrm{in}$ edges connecting it to members of the
same group and $z_\mathrm{out}$ edges to members of other groups, with
$z_\mathrm{in}$ and $z_\mathrm{out}$ chosen such that the total expected
degree $z_\mathrm{in}+z_\mathrm{out}=16$, in this case.  As
$z_\mathrm{out}$ is increased from small values, the resulting graphs pose
greater and greater challenges to the community-finding algorithm.  In
Fig.~\ref{correct} we show the fraction of vertices correctly assigned to
the four communities by the algorithm as a function of~$z_\mathrm{out}$.
As the figure shows, the algorithm performs well, correctly identifying
more than 90\% of vertices for values of $z_\mathrm{out}\lesssim6$.  Only
when $z_\mathrm{out}$ approaches the value~8 at which the number of within-
and between-community edges per vertex is the same does the algorithm begin
to fail.  On the same plot we also show the performance of the GN algorithm
and, as we can see, that algorithm performs slightly but measurably better
for smaller values of~$z_\mathrm{out}$.  For example, for
$z_\mathrm{out}=5$ our new algorithm correctly identifies an average of
$97.4(2)\%$ of vertices, while the older algorithm correctly identifies
$98.9(1)\%$.  Both, however, clearly perform well.

Interestingly for higher values of $z_\mathrm{out}$ our new algorithm
performs better than the older one, and we have come across a few
real-world networks in which this is the case also.  Normally, however, the
GN algorithm seems to have the edge, and this should come as no great
surprise.  Our new algorithm bases its decisions on purely local
information about individual communities, while the GN algorithm uses
non-local information about the entire network---information derived from
betweenness scores.  Since community structure is itself fundamentally a
non-local quantity, it seems reasonable that one can do a better job of
finding that structure if one has non-local information at one's disposal.

For systems small enough that the GN algorithm is computationally
tractable, therefore, we see no reason not to continue using it---it
appears to give the best results.  For systems too large to make use of
this approach, however, our new algorithm gives useful community structure
information with comparatively little effort.

\begin{figure}
\begin{center}
\resizebox{\figurewidth}{!}{\includegraphics{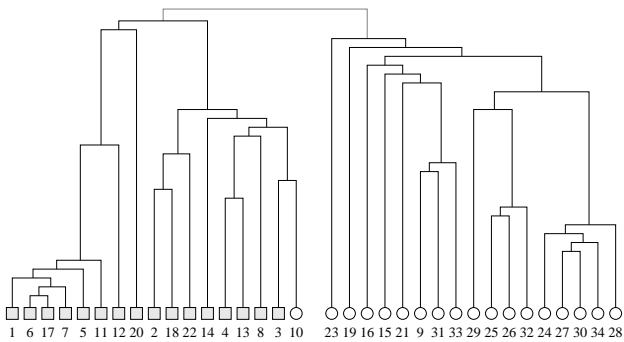}}
\end{center}
\caption{Dendrogram of the communities found by our algorithm in the
``karate club'' network of Zachary~\cite{Zachary77,GN02}.  The shapes of
the vertices represent the two groups into which the club split as the
result of an internal dispute.}
\label{zachary}
\end{figure}

We have applied our algorithm to a variety of real-world networks also.  We
have looked, for example, at the ``karate club'' network studied
in~\cite{GN02}, which represents friendships between 34 members of a club
at a US university, as recorded over a two-year period by
Zachary~\cite{Zachary77}.  During the course of the study, the club split
into two groups as a result of a dispute within the organization, and the
members of one group left to start their own club.  In Fig.~\ref{zachary}
we show the dendrogram derived by feeding the friendship network into our
algorithm.  The peak modularity is $Q=0.381$ and corresponds to a split
into two groups of~17, as shown in the figure.  The shapes of the vertices
represent the alignments of the club members following the split and, as we
can see, the division found by the algorithm corresponds almost perfectly
to these alignments; only one vertex, number~10, is classified wrongly.
The GN algorithm performs similarly on this task, but not better---it also
finds the split but classifies one vertex wrongly (although a different
one, vertex~3).  In other tests, we find that our algorithm also
successfully detects the main two-way division of the dolphin social
network of Lusseau~\cite{NG04,Lusseau03b}, and the division between black
and white musicians in the jazz network of Gleiser and Danon~\cite{GD04}.

\begin{figure*}[t]
\begin{center}
\resizebox{\textwidth}{!}{\includegraphics{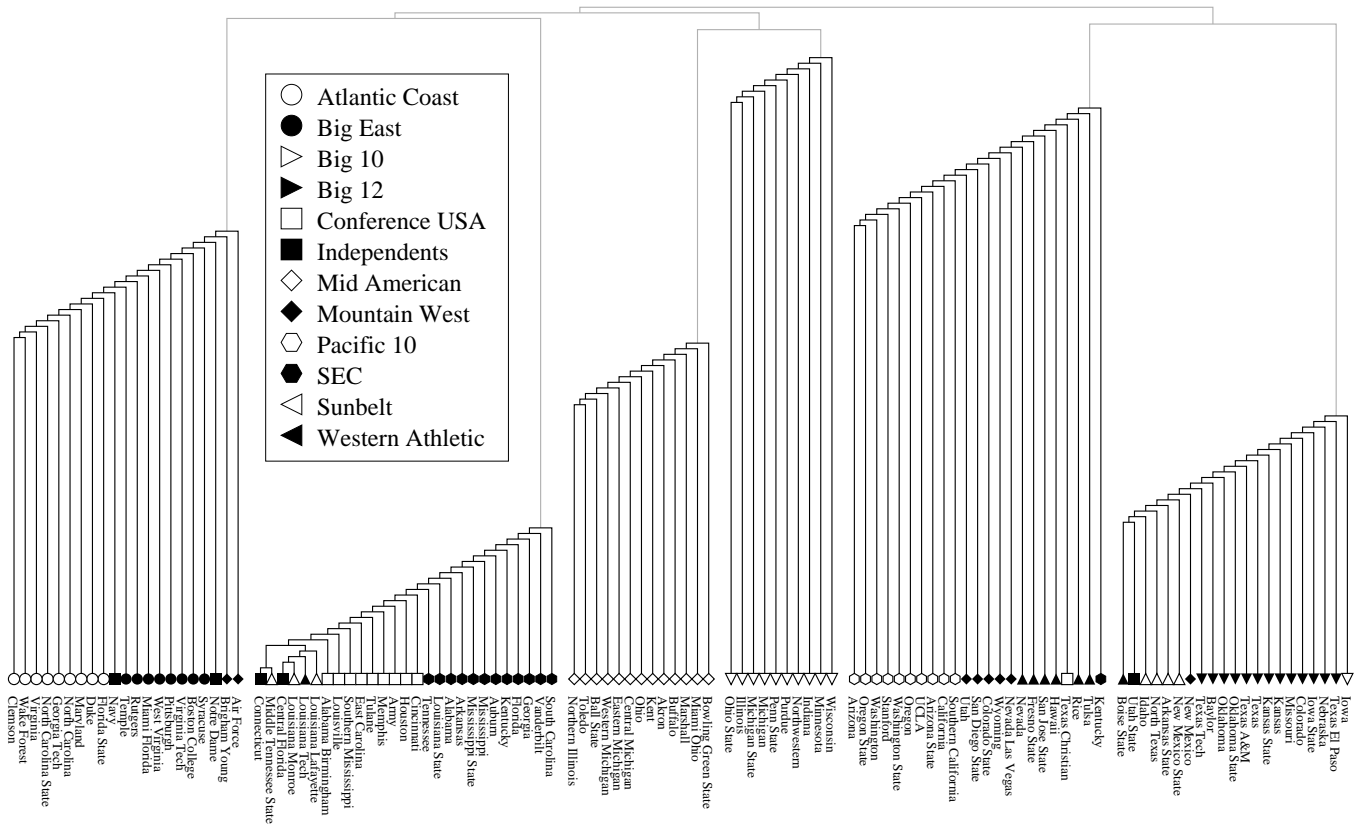}}
\end{center}
\caption{Dendrogram of the communities found in the college football
network described in the text.  The real-world
communities---conferences---are denoted by the different shapes as
indicated in the legend.}
\label{football}
\end{figure*}

As a demonstration of how our algorithm can sometimes miss some of the
structure in a network, we take another example from
Ref.~\onlinecite{GN02}, a network representing the schedule of games
between American college football teams in a single season.  Because the
teams are divided into groups or ``conferences,'' with intra-conference
games being more frequent than inter-conference games, we have a reasonable
idea ahead of time about what communities our algorithm should find.  The
dendrogram generated by the algorithm is shown in Fig.~\ref{football}, and
has an optimal modularity of $Q=0.546$, which is a little shy of the value
$0.601$ for the best split reported in~\cite{GN02}.  As the dendrogram
reveals, the algorithm finds six communities.  Some of them correspond to
single conferences, but most correspond to two or more.  The GN algorithm,
by contrast, finds all eleven conferences, as well as accurately
identifying independent teams that belong to no conference.  Nonetheless,
it is clear that the new algorithm is quite capable of picking out useful
community structure from the network, and of course it is much the faster
algorithm.  On the author's personal computer the algorithm ran to
completion in an unmeasureably small time---less than a hundredth of a
second.  The algorithm of Girvan and Newman took a little over a second.

A time difference of this magnitude will not present a big problem in most
practical situations, but performance rapidly becomes an issue when we look
at larger networks; we expect the ratio of running times to increase with
the number of vertices.  Thus, for example, in applying our algorithm to
the 1275-node network of jazz musician collaborations mentioned above, we
found that it runs to completion in about one second of CPU time.  The GN
algorithm by contrast takes more than three hours to reach very similar
results.

\begin{figure*}[t]
\begin{center}
\resizebox{\textwidth}{!}{\includegraphics{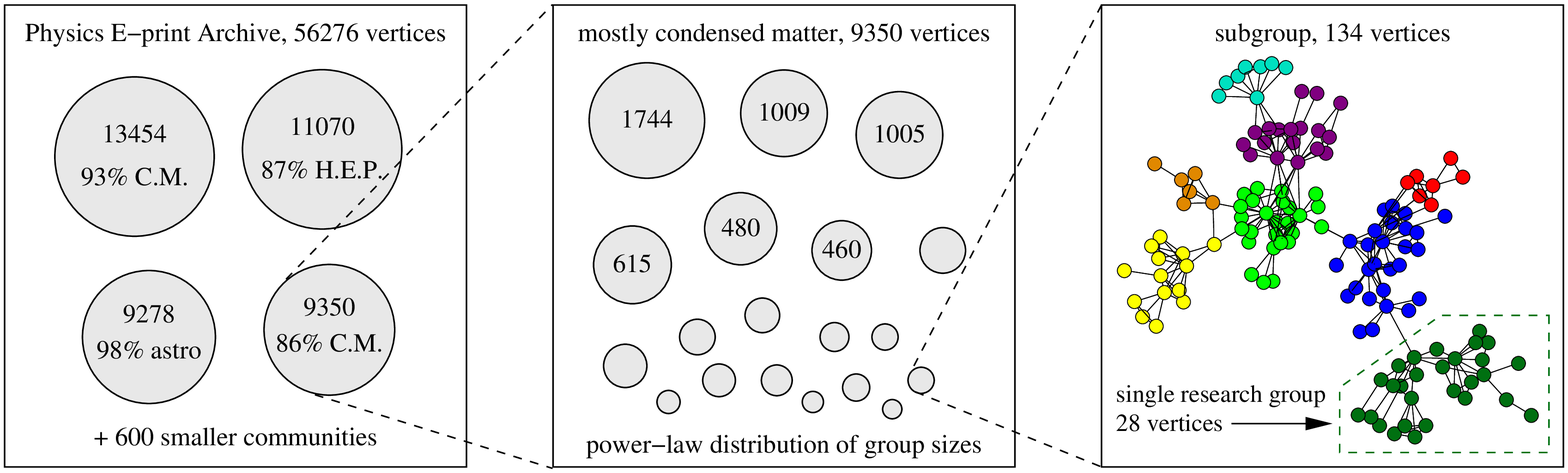}}
\end{center}
\caption{Left panel: Community structure in the collaboration network of
physicists.  The graph breaks down into four large groups, each composed
primarily to physicists of one specialty, as shown.  Specialties are
determined by the subsection(s) of the e-print archive in which individuals
post papers: ``C.M.'' indicates condensed matter; ``H.E.P.'' high-energy
physics including theory, phenomenology, and nuclear physics; ``astro''
indicates astrophysics.  Middle panel: one of the condensed matter
communities is further broken down by the algorithm, revealing an
approximate power-law distribution of community sizes.  Right panel: one of
these smaller communities is further analyzed to reveal individual research
groups (different shades), one of which (in dashed box) is the author's
own.}
\label{arxiv}
\end{figure*}

As an example of an analysis made possible by the speed of the new
algorithm, we have looked at a network of collaborations between physicists
as documented by papers posted on the widely-used Physics E-print Archive
at \texttt{arxiv.org}.  The network is an updated version of the one
described in Ref.~\onlinecite{Newman01a}, in which scientists are
considered connected if they have coauthored one or more papers posted on
the archive.  We analyze only the largest component of the network, which
contains $n=56\,276$ scientists in all branches of physics covered by the
archive.  Since two vertices that are unconnected by any path are never put
in the same community by our algorithm, the small fraction of vertices that
are not part of the largest component can safely be assumed to be in
separate communities in the sense of our algorithm.  Our algorithm takes 42
minutes to find the full community structure.  Our best estimates indicate
that the GN algorithm would take somewhere between three and five years to
complete its version of the same calculation.

The analysis reveals that the network in question consists of about 600
communities, with a high peak modularity of $Q=0.713$, indicating strong
community structure in the physics world.  Four of the communities found
are large, containing between them 77\% of all the vertices, while the
others are small---see Fig.~\ref{arxiv}, left panel.  The four large
communities correspond closely to subject subareas: one to astrophysics,
one to high-energy physics, and two to condensed matter physics.  Thus
there appears to be a strong correlation between the structure found by our
algorithm and the community divisions perceived by human observers.  It is
precisely correlation of this kind that makes community structure analysis
a useful tool in understanding the behavior of networked systems.

We can repeat the analysis with any of the subcommunities to observe how
they break up.  For example, feeding the smaller of the two
condensed-matter groups through the algorithm again, we find an even
stronger peak modularity of $Q=0.807$---the strongest we have yet observed
in any network---corresponding to a split into about a hundred communities
of all sizes (Fig.~\ref{arxiv}, center panel).  The sizes appear roughly to
have a power-law distribution with exponent about~$-2$~\footnote{This power
law is of a different kind to the one observed in an email network by
Guimer\`a~\etal~\cite{Guimera03}.  They studied the histogram of community
sizes over all levels of the dendrogram; we are looking only at the single
level corresponding to the maximum value of~$Q$.}.  Narrowing our focus
still further to the particular one of these communities that contains the
present author, we find the structure shown in the right panel of
Fig.~\ref{arxiv}.  Feeding this one last time through the algorithm, it
breaks apart into communities that correspond closely to individual
institutional research groups, the author's group appearing in the corner
of the figure, highlighted by the dashed box.  One could pursue this line
of analysis further, identifying individual groups, iteratively breaking
them down, and looking for example at the patterns of collaboration between
them, but we leave this for later studies.

\section{Conclusions}
In this paper we have described a new algorithm for extracting community
structure from networks, which has a considerable speed advantage over
previous algorithms, running to completion in time that scales as the
square of the network size.  This allows us to study much larger systems
than has previously been possible.  Among other examples, we have applied
the algorithm to a network of collaborations between more than fifty
thousand physicists, and found that the resulting community structure
corresponds closely to the traditional divisions between specialties and
research groups in the field.

We believe that our method will not only allow for the extension of
community structure analysis to some of the very large networks that are
now being studied for the first time, but also provides a useful tool for
visualizing and understand the structure of these networks, whose
daunting size has hitherto made many of their structural properties
obscure.

\begin{acknowledgments}
The author thanks Leon Danon, Pablo Gleiser, David Lusseau, and Douglas
White for providing network data used in the examples.  This work was
supported in part by the National Science Foundation under grant number
DMS--0234188.
\end{acknowledgments}

\end{document}